\documentclass[a4paper]{jpconf}
\usepackage{graphicx}
\usepackage{amsmath}
\begin{document}

\title{
	\begin{flushright}
		\ \\*[-80pt] 
		\begin{minipage}{0.2\linewidth}
			\normalsize
			HUPD-2211 \\*[5pt]
		\end{minipage}
	\end{flushright}
Study of weak-basis invariants in the universal seesaw model using Hilbert series}

\author{Albertus Hariwangsa Panuluh$^{1,2}$ and Takuya Morozumi$^{1,3}$}

\address{$^1$Physics Program, Graduate School of Advanced Science and Engineering, Hiroshima University, Higashi-Hiroshima 739-8526, Japan} \address{$^2$Department of Physics Education, Sanata Dharma University, Paingan, Maguwohardjo, Sleman, Yogyakarta 55282, Indonesia}
\address{$^3$Core of Research for Energetic Universe, Hiroshima University, Higashi-Hiroshima 739-8526, Japan}

\ead{panuluh-albertus@hiroshima-u.ac.jp, morozumi@hiroshima-u.ac.jp}

\begin{abstract}
Universal Seesaw Model is a model which explains the mass hierarchy of the quark sector. This model introduces vector-like quarks. The top quark mass is generated in the electroweak scale and the other quark mass is generated using a seesaw-like mechanism. The invariant theory helps construct a weak-basis invariant. We study the weak-basis invariant (WBI) using Hilbert Series (HS) and apply it to the Universal Seesaw Model, particularly the one-generation case of the quark sector. 

\end{abstract}

\section{Introduction}
Standard Model (SM) is the most successful theory that incorporates the dynamics of a sub-atomic particle and its CP symmetry is broken by complex phase in the Yukawa couplings of quark sector\cite{kobayashi}. However, SM cannot explain some phenomena, one of which is the fermion mass hierarchy. Universal seesaw model is one model that aims to explain this problem\cite{ber,raj,chang,dav}. One has studied the quark sector CP violation of the universal seesaw model\cite{ume}, and they found that even in one generation case, there is an imaginary parameter that leads to CP violation. Moreover, the flavor structures are also more complicated than SM. These flavor structures are completely arbitrary and basis dependent. So one can apply weak-basis transformation (WBT) on the fields and the physical observables have to be independent under this change of basis. These quantities are called weak-basis invariants (WBI)\cite{jarls,branco}. 

We aim to study the WBI in the universal seesaw model, particularly the one-generation case of the quark sector. To find the WBI in this model, we use the invariant theory method so-called Hilbert series (HS)\cite{man}.

\section{The model}
The gauge group of the model is $SU(3)_C \times SU(2)_L \times SU(2)_R \times U(1)'_Y$. The complete particle contents and their charge assignments under the underlying gauge group are given in Table \ref{model}. After spontaneously symmetry breaking (SSB),  the ordinary quark and VL quarks will mix at the tree level from non-zero vacuum expectation value (VEV) of $SU(2)_L$ and $SU(2)_R$ doublet Higgs $v_L$ and $v_R$ respectively, which satisfy $v_R\gg v_L$. 

\begin{table}[h]
	\caption{\label{model}Quark and higgs field in the model and their charge assignments.} 
	
	\begin{center}
		\lineup
		\begin{tabular}{ccccc}
			\hline
			Quark and Higgs Field	& $SU(3)_C$ & $SU(2)_L$ &$SU(2)_R$  & $U(1)_{Y'}$ \\
			\hline\hline
			$q_L=\left( \begin{array}{c}
				u_L	\\d_L
			\end{array}\right)$ 	& \textbf{3} & \textbf{2} & \textbf{1} & 1/6 \\
			
			$q_R=\left( \begin{array}{c}
				u_R	\\d_R
			\end{array}\right)$	& \textbf{3} &  \textbf{1}& \textbf{2} & 1/6 \\
			
			$U_{L,R}$	& \textbf{3} & \textbf{1} &  \textbf{1}& 2/3 \\
			
			$D_{L,R}$	& \textbf{3} &\textbf{1}  & \textbf{1} & $-1/3$ \\
			
			$\phi_L=\left( \begin{array}{c}
				\phi_L^+	\\\phi_L^0
			\end{array}\right)$ 	& \textbf{1} &\textbf{2}  &\textbf{1}  &1/2  \\
			
			$\phi_R=\left( \begin{array}{c}
				\phi_R^+	\\\phi_R^0
			\end{array}\right)$	&\textbf{1}  & \textbf{1} &\textbf{2}  &  1/2\\
			\hline
		\end{tabular}
	\end{center}
\end{table}

Considering one generation case without flavor mixing, the Yukawa interaction and the mass terms of VL-quark as follows,
\begin{align}\label{lagrangian}
	\mathcal{L}=&-\overline{q_L}y_{uL}\tilde{\phi}_L U_{R}-\overline{q_{R}}y_{uR}\tilde{\phi}_R U_{L}-\overline{U_{L}}M_U U_{R}-h.c.\nonumber\\
	&- \overline{q_{L}}y_{dL}{\phi}_L D_{R}-\overline{q_{R}}y_{dR}{\phi}_R D_{L}-\overline{D_{L}}M_DD_{R}-h.c.
\end{align}
where $y_{uL},y_{uR},y_{dL},y_{dR}$ are Yukawa couplings and $M_{U(D)}$ is the up-type (down-type) VL-quark mass. 
For instance, after substituting the non-zero vev of L-R Higgs, changing to mass eigenstate basis, and diagonalizing the mass matrix using bi-unitary transformation, the approximate light and heavy mass eigenvalues for up(down)-type quark are,
\begin{equation}
	m_{u(d)_1}\simeq \frac{v_L v_R |y_{u(d)L}| |y_{u(d)R}|}{2 M_{U(D)}},\qquad m_{u(d)_2}\simeq M_{U(D)}
\end{equation}
respectively.

\section{Invariant theory}
In this section, we give explanation about invariant theory briefly where mainly we studied from \cite{fi}. Consider a model containing $n$ parameters $q=(q_1,q_2,\ldots,q_n)^T$ and are transformed under a symmetry group $G$ representation $R(g)$, where $g \in G$, as follows,
\begin{equation}
	q^\prime=R(g)q,\qquad \forall g \in G.
\end{equation}
Invariants denoted by $I(q)$ are polynomial functions of $q$ and satisfy $I(q')=I(q)$ and they form a ring (due to the addition and multiplication property). The ring will have a finite number of generators (basic invariants) if $G$ is a reductive group\cite{man}. However, there might be non-trivial relations among the basic invariants. The number of linearly independent (primary invariants) is equal to the number of physical parameters in the model. These invariants can be obtained using Hilbert series\cite{man},
\begin{eqnarray} \label{hs}
	\mathcal{H}(q)=\sum_{r=0}^{\infty}c_r q^r=\frac{1+c_1q+\cdots+c_{k-1}q^{k-1}+q^k}{\prod_{r=1}^{p}(1-q^{d_r})}.
\end{eqnarray}
\begin{eqnarray*}
	&&q : |q|<1 \hspace{3pt} \text{complex number labeling the building blocks} \nonumber\\
	&&c_r : \text{number of invariants of degree} \hspace{3pt} r \nonumber \\
	&&k : \text{degree of numerator}; \qquad p : \text{number of parameters}
\end{eqnarray*}
\section{Weak-basis invariants (WBI)}
Define the WBT on doublet and singlet VL quarks as follows,
\begin{align}
	&q_L'=e^{i\theta_{V_L}}q_L,\qquad 	q_R'=e^{i\theta_{V_R}}q_R,\qquad U_L'=e^{i\theta_{U_L}}U_L \nonumber\\
	&U_R'=e^{i\theta_{U_R}}U_R,\qquad D_L'=e^{i\theta_{D_L}}D_L, \qquad D_R'=e^{i\theta_{D_R}}D_R.
\end{align}
The lagrangian in Eq.(\ref{lagrangian}) unchanged if the Yukawa couplings and VL quark mass are transformed,

\begin{align}
	&y_{uL}'=e^{i\theta_{V_L}}y_{uL}e^{-i\theta_{U_R}}, \quad (y_{uL}^{\ast})'=e^{i\theta_{U_R}}y_{uL}^\ast e^{-i\theta_{V_L}}, \quad y_{uR}'=e^{i\theta_{V_R}}y_{uR}e^{-i\theta_{U_L}} \nonumber\\ &(y_{uR}^{\ast})'=e^{i\theta_{U_L}}y_{uR}^\ast e^{-i\theta_{V_R}}, \quad
	y_{dL}'=e^{i\theta_{V_L}}y_{uL}e^{-i\theta_{D_R}}, \quad (y_{dL}^{\ast})'=e^{i\theta_{D_R}}y_{dL}^\ast e^{-i\theta_{V_L}} \nonumber\\ &y_{dR}'=e^{i\theta_{V_R}}y_{dR}e^{-i\theta_{D_L}}, \quad (y_{dR}^{\ast})'=e^{i\theta_{D_L}}y_{dR}^\ast e^{-i\theta_{V_R}},\quad
	M_U'=e^{i\theta_{U_L}}M_U e^{-i\theta_{U_R}}\nonumber\\ &(M_U^\ast)'=e^{i\theta_{U_R}}M_U^\ast e^{-i\theta_{U_L}}, \quad M_D'=e^{i\theta_{D_L}}M_D e^{-i\theta_{D_R}}, \quad (M_D^\ast)'=e^{i\theta_{D_R}}M_D^\ast e^{-i\theta_{D_L}}.
\end{align}
By labeling $q_1=y_{uL},q_2=y_{uL}^\ast,q_3=y_{uR},q_4=y_{uR}^\ast,q_5=y_{dL},q_6=y_{dL}^\ast,q_7=y_{dR},q_8=y_{dR}^\ast,q_9=M_U,q_{10}=M_U^\ast,q_{11}=M_D,q_{12}=M_D^\ast$ and by using Molien function\cite{book} (or also known as Molien-Weyl formula\cite{fi}), we obtain the multi-graded HS,
\begin{align}\label{multi}
	H(q_1,\ldots, q_{12})=&\frac{1-q_1 q_2 q_3 q_4 q_5 q_6 q_7 q_8 q_9 q_{10} q_{11} q_{12}}{(1-q_1 q_2)(1-q_3 q_4)(1-q_5 q_6)(1-q_7 q_8)(1-q_9 q_{10})}\nonumber\\&\times \frac{1}{(1-q_{11} q_{12})(1-q_2 q_3 q_5 q_8 q_9 q_{12}) (1-q_1 q_4 q_6 q_7 q_{10} q_{11})}.
\end{align}
From the denominator of Eq.(\ref{multi}), we obtained information about basic WBIs. We have eight basic WBIs denoted as follow: 
\begin{align}
	I_1&=y_{uL}y_{uL}^\ast,\quad I_2=y_{uR}y_{uR}^\ast, \quad I_3=y_{dL}y_{dL}^\ast, \quad I_4=y_{dR}y_{dR}^\ast,\quad I_5=M_U M_U^\ast \nonumber\\ I_6&=M_D M_D^\ast, \quad I_7=y_{uL}^\ast y_{uR} y_{dL}y_{dR}^\ast M_U M_D^\ast, \quad I_8 = y_{uL}y_{uR}^\ast y_{dL}^\ast y_{dR} M_U^\ast M_D. 
\end{align}
In addition, from the numerator, we obtain one relation between basic WBIs $I_7 I_8 = I_1 I_2 I_3 I_4 I_5 I_6$. Using $\mathcal{H}(q)=H(q,q,\ldots,q)$ relation, where $\mathcal{H}(q)$ is ungraded HS (eq.(\ref{hs})) and $H(q,q,\ldots,q)$ is Eq.(\ref{multi}) with substituting all variables $q_i=q$ where $i=1,\ldots,12$, we get the ungraded HS in this model as follows,
\begin{eqnarray}\label{ungrad}
	\mathcal{H}(q)=\frac{1+q^6}{(1-q^2)^6 (1-q^6)}.
\end{eqnarray} 
From the denominator of Eq.(\ref{ungrad}), we see that there are six primary WBIs of degree two, which are $I_1\sim I_6$ and one primary WBI of degree six which we choose as $I_7-I_8\equiv J_9$. The other degree six invariant ($I_7+I_8\equiv J_{10}$) is counted in the numerator. Therefore, we have seven primary WBIs  which means there are seven parameters in this one-generation case of the quark sector universal seesaw model.

Moreover, we also looked at the CP parities of the primary WBIs. We have six CP even ($I_1 \sim I_6$) and one CP odd ($J_9$). From \cite{ume}, there is one CP violating WBI in one-generation case of the quark sector universal seesaw model. We write the CP violating WBI in terms of invariants in our results as,
\begin{equation}
	W=-\frac{J_9}{2i}
\end{equation}
which appears in the $W_L-W_R$ mixing \cite{yam}.
\section{Summary}
We have implemented Hilbert Series method in the universal seesaw model, particularly to the one generation case of quark sector. We find eight basic WBIs becoming the generator of the ring, one syzygy and seven primary WBIs. Furthermore, there is one CP violating WBI in this model which appears in the $W_L-W_R$ mixing.
\section*{Acknowledgment}
We would like to thank Prof. Taku Yamanaka as chair of the International Conference on Kaon Physics (KAON 2022) and all the organizers' members for organizing such a wonderful conference and for their hospitality. We also thank K. Yamamoto for the fruitful discussion. A.H.P would like to thank the Ministry of Education, Culture, Sports, Science, and Technology, Japan (MEXT) for the financial support during this work.
The work of T. M. is supported by Japan Society for the Promotion of Science
(JSPS) KAKENHI Grant Number JP17K05418.

\end{document}